\documentclass[twocolumn,floatfix,altaffilletter,superscriptaddress,preprintnumbers,
               tightenlines,showpacs,showkeys,nofootinbib]{revtex4-1}
\usepackage[utf8]{inputenc}
\usepackage[colorlinks=true,citecolor=blue,linkcolor=blue]{hyperref}
\usepackage[normalem]{ulem}
\usepackage{amsmath,amssymb, mathrsfs}
\usepackage{epsfig}
\usepackage{graphicx}               
\usepackage{url}
\usepackage{color}
\usepackage{slashed}
\usepackage{multirow}
\usepackage{placeins}
\usepackage[dvipsnames]{xcolor}
\usepackage{epstopdf}
\usepackage{soul}
\usepackage{tikz}
\usepackage[capitalise, english]{cleveref}
\usepackage{siunitx}
\usepackage{xspace}
\usepackage{booktabs}
\usetikzlibrary{trees}
\usetikzlibrary{decorations.pathmorphing}
\usetikzlibrary{decorations.markings}

\newcommand\myshade{80}
\colorlet{mylinkcolor}{ForestGreen}
\colorlet{mycitecolor}{Red}
\colorlet{myurlcolor}{violet}

\hypersetup{
  linkcolor  = mylinkcolor!\myshade!black,
  citecolor  = mycitecolor!\myshade!black,
  urlcolor   = myurlcolor!\myshade!black,
  colorlinks = true
}

\definecolor{jblue}{RGB}{20,50,100}
\definecolor{npurple}{RGB} {153, 51, 204}
\definecolor{wred}{RGB}{217,0,56}
\definecolor{white}{RGB}{255,255,255}

\definecolor{korange}{RGB}{235, 80,  43}
\definecolor{korange2}{RGB}{245, 100,  63}
\definecolor{kyelloworange}{RGB}{255, 210,  110}
\definecolor{kyelloworange2}{RGB}{240, 170,  90}
\definecolor{kred}{RGB}{204,  102, 153}
\definecolor{kpurple}{RGB}{153,  61, 190}
\definecolor{kpurplelight}{RGB}{213,  161, 230}

 \definecolor{tobycolour}{rgb}{.5,.0,.5}

\DeclareSIUnit\year{yr}
\DeclareSIUnit\pc{pc}
\DeclareSIUnit\ergs{ergs}
\DeclareSIUnit\msun{\ensuremath{M_\odot}}
\allowdisplaybreaks

\newcommand{\lmultau}{\ensuremath{L_\mu-L_\tau}\xspace}
\newcommand{\mv}{\ensuremath{m_V}\xspace}
\newcommand{\gv}{\ensuremath{g^\prime}\xspace}

\makeatletter
\providecommand*{\diff}%
  {\@ifnextchar^{\DIfF}{\DIfF^{}}}
\def\DIfF^#1{%
  \mathop{\mathrm{\mathstrut d}}%
    \nolimits^{#1}\gobblespace}
\def\gobblespace{%
  \futurelet\diffarg\opspace}
\def\opspace{%
  \let\DiffSpace\!%
  \ifx\diffarg(%
    \let\DiffSpace\relax
  \else
    \ifx\diffarg[%
      \let\DiffSpace\relax
    \else
        \ifx\diffarg\{%
        \let\DiffSpace\relax
      \fi\fi\fi\DiffSpace}


\begin{document}

\title{Probing muonic forces with neutron star binaries}

\author{Jeff A. Dror}
\email{jdror@lbl.gov 0000-0003-0110-6184}
\affiliation{Theory Group, Lawrence Berkeley National Laboratory, Berkeley, CA 94720, USA}
\affiliation{Berkeley Center for Theoretical Physics, University of California, Berkeley, CA 94720, USA}

\author{Ranjan Laha} 
\email{ranjan.laha@cern.ch 0000-0001-7104-5730}
\author{Toby Opferkuch}
\email{toby.opferkuch@cern.ch 0000-0002-7388-7453}
\affiliation{Theoretical Physics Department, CERN, 1211 Geneva, Switzerland}

\date{\today}

\preprint{CERN-TH-2019-150}

\begin{abstract}
We show that gravitational wave emission from neutron star binaries can be used to discover any generic long-ranged muonic force due to the large inevitable abundance of muons inside neutron stars. As a minimal consistent example, we focus on a gauged U(1)$_{\lmultau}$ symmetry. In pulsar binaries, such U(1)$_{\lmultau}$ vectors induce an anomalously fast decay of the orbital period through the emission of dipole radiation. We study a range of different pulsar binaries, finding the most powerful constraints for vector masses below ${\cal O}(\SI{E-18}{\eV})$. For merging binaries the presence of muons in neutron stars can result in dipole radiation as well as a modification of the chirp mass during the inspiral phase. We make projections for a prospective search using both the GW170817 and S190814bv events and find that current data can discover light vectors with masses below ${\cal O}(\SI{E-10}{\eV})$. In both cases, the limits attainable with neutron stars reach gauge coupling $g^\prime\lesssim 10^{-20}$, which are many orders of magnitude stronger than previous constraints. We also show projections for next generation experiments, such as Einstein Telescope and Cosmic Explorer.
\end{abstract}

\maketitle

\section{Introduction}

New long range interactions give rise to distinctive signatures in a wide range of observables. Such interactions are, however, strongly constrained by fifth-force tests~\cite{Adelberger:2003zx, Murata:2014nra,Adelberger:2006dh,Schlamminger:2007ht}, unless they are either screened or the couplings to the first generation fermions are suppressed. We show that the dynamics of neutron star (NS) binaries provide ideal laboratories to probe long range muonic forces due to the significant abundance of muons inside them (by mass $\gtrsim$ 0.1\% M$_\odot$). The observations of the NS merger event, GW170817~\cite{TheLIGOScientific:2017qsa}, an NS-black hole (BH) candidate merger event, S190814bv~\cite{LIGO_BHNS,Ackley:2020qkz}, and various pulsar binaries \cite{Kramer:2006nb,vanStraten:2001zk,Stairs:2002cw,Shannon2013,Antoniadis:2013pzd,Bhat:2008ck,Freire:2012mg,Ferdman:2014rna,vanLeeuwen:2014sca,Jacoby:2006dy} give us the opportunity to probe these new exotic forces. These methods of probing muonic forces via NS binaries are completely general, applicable to both vector and scalar mediators. However, as a concrete realization we focus on a long-range gauged U(1) symmetry. 

Additional U(1) gauge symmetries with masses below the weak scale are simple extensions of Standard Model that can act as a mediator to the dark sector (see, e.g., \cite{Boehm:2003hm,Fayet:2006sp,Pospelov:2007mp,ArkaniHamed:2008qn}), are common predictions of string theory~\cite{Goodsell:2009xc}, and can explain experimental anomalies~\cite{Gninenko:2001hx,Kahn:2007ru,TuckerSmith:2010ra,Batell:2011qq,Feng:2016ysn,Altmannshofer:2014cfa}. The observed matter content limits the linearly independent conserved currents to $B-L$, hypercharge (equivalent to kinetic mixing), and (up to neutrino masses) $ L_e-L_\mu$, $ L_\mu-L_\tau$, and $L_e-L_\tau $.\footnote{Relaxing the requirement of anomaly cancellation with just the Standard Model fields greatly enlarges the possibilities but is highly constrained from searches for flavor changing neutral currents~\cite{Dror:2017ehi,Dror:2017nsg,Laha:2013xua}.} The small number of possibilities highlights the need to find all experimental ways to probe these light vectors. While most of the focus when studying the phenomenology of new U(1) gauge symmetries has been above the {\rm eV} scale, it is interesting to study the constraints for lighter masses, where forces are long-ranged. For $ B-L $, $L_e-L_\mu$, and $L_e-L_\tau$, the constraints below this scale become extremely powerful from fifth force tests~\cite{Adelberger:2003zx, Murata:2014nra,Adelberger:2006dh,Schlamminger:2007ht} which constrains such forces to be weaker than gravity once the vector mass drops below ${\cal O}(\SI{E-4}{\eV})$ and constraining the gauge coupling $g^\prime\lesssim \SI{E-20}{}$ at the lowest masses. Kinetically mixed dark photons do not experience such constraints due to screening of the charge between protons and electrons leading to rich phenomenology at low masses (see, e.g., \cite{Arias:2012az} and references therein). Interestingly, \lmultau forces also are not bound by these constraints since the muon fraction in ordinary matter is negligible.

\begin{figure}	
  \centering
  \includegraphics[width=\columnwidth]{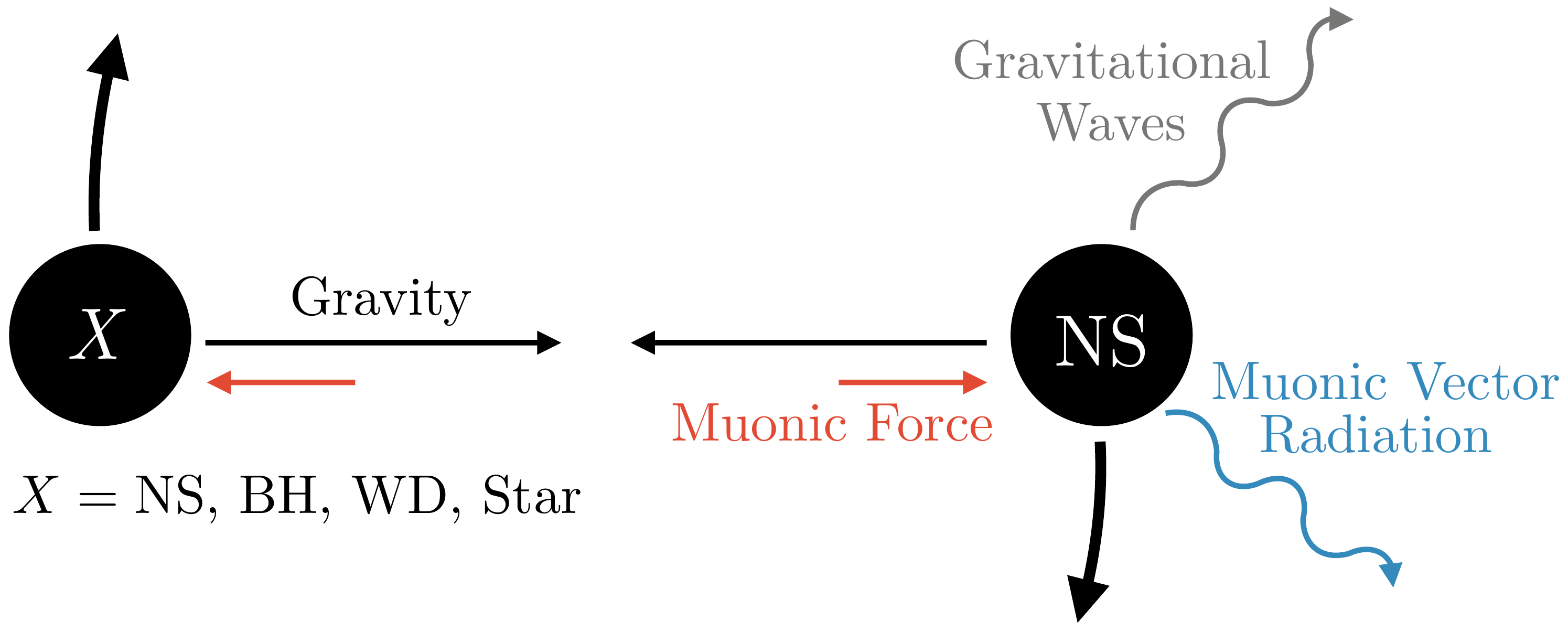}
  \caption{Sketch of an NS binary system showing the modifications induced by a generic muonic repulsive force and the radiation of the force mediator. Note that the force between the bodies is only present for the case of an NS--NS binary, i.e., $X=\text{NS}$.}
  \label{fig:NS_inspiral_sketch}
\end{figure}

\begin{figure*}
  \centering
  \includegraphics{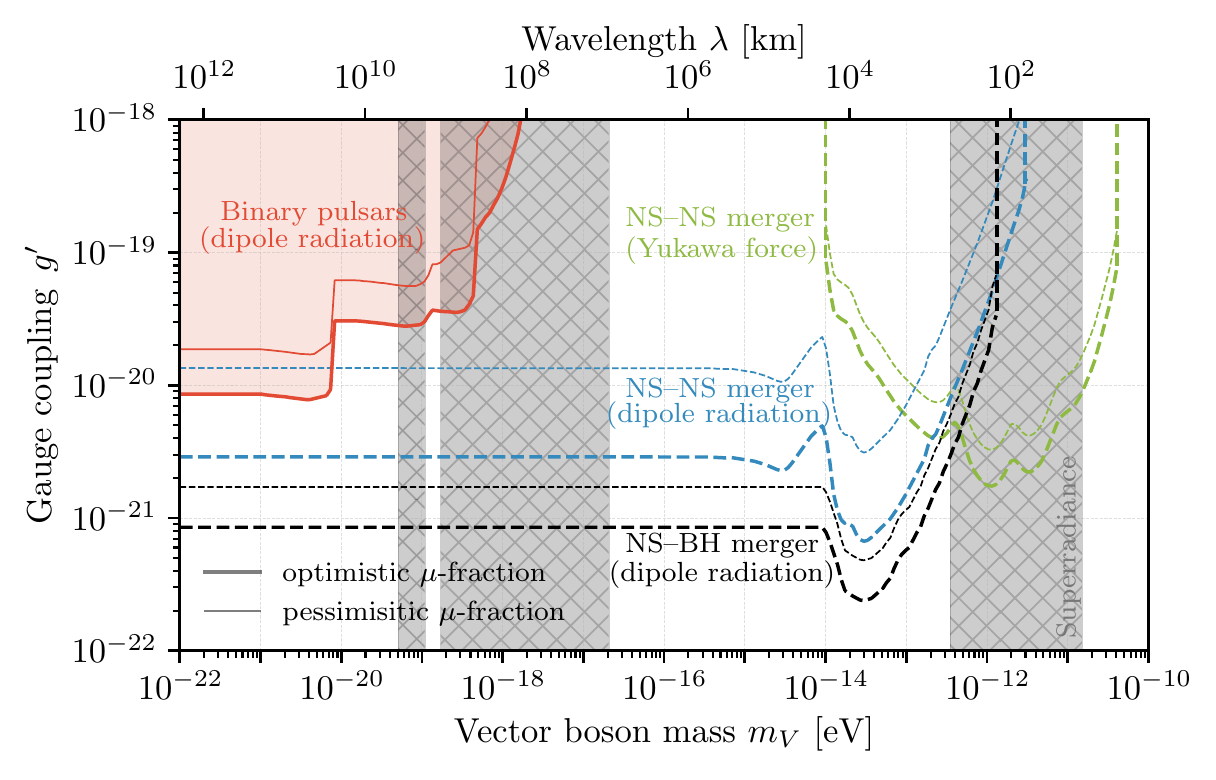}
  \caption{Current sensitivity of NS binaries to a $ L _\mu - L _\tau $ gauge coupling, \gv, as a function of the vector mass, \mv. The merger curves are projections and require a dedicated analysis to be carried out by the LIGO Collaboration. The gray hatched regions indicate parameter space where the light vector is constrained by BH superradiance considerations~\cite{Baryakhtar:2017ngi}. See \cref{sec:discussion} for the discussion of the boundaries of these constraints.}
  \label{fig:mV_vs_gV_constraints}
\end{figure*}

For vector masses above an MeV, the strongest constraints on \lmultau range from beam dump experiments, muonic $g-2$ measurements, neutrino trident processes, and collider experiments (see e.g.,~\cite{Escudero:2019gzq,Altmannshofer:2014pba,Altmannshofer:2014cfa,Kamada:2018zxi,Asai:2018ocx,Foldenauer:2018zrz,Arcadi:2018tly,Bauer:2018egk,Altmannshofer:2016jzy,Biswas:2016yan,Heeck:2010pg,Heeck:2011wj,Alimena:2019zri,Garani:2019fpa,Joshipura:2019qxz,Krnjaic:2019rsv,Galon:2019owl}). For lower vector masses, the best published constraints on \lmultau arise from $\Delta N_\text{eff}$ through observations of Big Bang Nucleosynthesis~\cite{Grifols:1996gn,Kamada:2015era}, from SN1987A~\cite{Gninenko:1997iy}\footnote{The robustness of the supernova bounds has recently been called into question in~\cite{Bar:2019ifz}.}, and neutrino self-interactions \cite{Kreisch:2019yzn,Smirnov:2019cae,Chu:2018gxk} constraining, $g^\prime\lesssim 10^{-5}$.~\footnote{If we insist on a reheating temperature above the muon mass then there is a stronger bound of $ g ' \lesssim 10 ^{ - 9} $\cite{Krnjaic:2019rsv}.} Such searches are inherently much weaker than searches for long range forces as they do not scale with the size of the apparatus, which will allow us to present constraints much stronger than in previous work.  Alternatively, if there exists a mass mixing between the Standard Model $Z$ boson and the new vector, ${\cal L}\supset\varepsilon_Z m_Z^2 X_\mu Z^\mu$, it can induce a long range force that would have been observed in neutrino oscillations unless $\varepsilon_Z g^\prime \lesssim 10 ^{-52}$~\cite{Heeck:2010pg}. However, since a mass-mixing can only arise after gauge symmetry breaking, it is naturally small and is highly constrained experimentally~\cite{Dror:2018wfl}.

The existence of NSs is contingent entirely on the stability of the neutron through sufficient Pauli blocking of the process $n\to p + e^- + \overline{\nu}_e$ \cite{PhysRev.140.B1445,PhysRev.140.B1452,Potekhin:2013qqa,Goriely:2010bm,PhysRevC.88.024308}. As the neutron density increases similar processes involving muons, rather than electrons, become energetically favorable. This leads to the production of a significant number of muons and forbids their subsequent decay~\cite{Cohen1970}. NSs with masses of order a solar mass subsequently have $0.15\%$--$0.75\%$ of their mass stored in muons, providing a unique laboratory to test couplings of muons to light new degrees of freedom. This has been leveraged to place constraints on muon-philic dark matter due to its accretion in NSs~\cite{Garani:2019fpa,Bell:2019pyc,Garani:2018kkd}.

A key feature of this muon population is their asymmetric nature, i.e., the production of only muons and not antimuons. Consequently, the presence of new long-range forces coupled to muons leads to NSs acquiring large effective charges. The coupling we are interested in constraining is that of \lmultau,
\begin{equation} 
{\cal L} \supset g' V _\alpha \left(  \bar{\mu} \gamma ^\alpha \mu -  \bar{ \tau } \gamma ^\alpha \tau + \bar{\nu} _\mu \gamma ^\alpha \nu _\mu - \bar{\nu} _\tau \gamma ^\alpha \nu _\tau \right) 
\end{equation} 
where $V_\alpha$ and $g^\prime$ denote the new vector and its coupling strength. We will rely on the couplings to muons as NSs contain negligible numbers of taus or neutrinos. Extending the constraints we derive here to a scalar force is straightforward, where $\mathcal{O}(1)$ changes are expected in the limits on the coupling relative to the vector case.  

 The observation of gravitational waves (GWs) from NS--NS/ NS--BH mergers as well as timing measurements of binary pulsars provides exquisite sensitivity to not only these scenarios, but also many types of dark matter candidates (see \cite{Bertone:2019irm} and references therein). In both cases, GWs are the dominant energy loss mechanism required to describe the dynamics of the system. For the case of mergers, sensitivity arises from the long observation duration of the post-Newtonian stage of the GW signal \cite{TheLIGOScientific:2017qsa}. While for the case of binary pulsars, precise measurements of the change in pulsar period yield accurate determinations of both the relativistic corrections to the  binary orbits and the orbital decay, (see, e.g.~\cite{Manchester:2015mda}). Therefore, both of these systems are sensitive to additional energy loss mechanisms through the emission of the light vector, while NS mergers are also sensitive to an additional force sourced between the muon content of the two NSs; see \cref{fig:NS_inspiral_sketch}. These effects have been studied in the context of new long-range forces in a hidden sector \cite{Croon:2017zcu,Sagunski:2017nzb,Kopp:2018jom,Alexander:2018qzg,Choi:2018axi,Fabbrichesi:2019ema,Hook:2017psm,Huang:2018pbu,Seymour:2019tir}. In what follows, we consider the impact a gauged U(1)$_{\lmultau}$ symmetry can have on NS binaries.

\section{Muons in neutron stars}
\label{sec:muons_in_NSs}

The presence of muons in NSs arises due to chemical equilibrium and charge neutrality maintained via the processes $n\to p + e^-/\mu^- + \overline{\nu}_{e/\mu}$ and $e^-\to \mu^- +\overline{\nu}_\mu+\nu_e$, referred to as beta equilibrium. The existence of muons in NSs follows from estimating the Fermi energy of a neutron gas, $ E_F = (3\pi^2 n_n)^{2/3}/2m_n $ (where $m_n$ and $n_n$ are the nucleon mass and number density of the neutrons, respectively). NS masses around a solar mass and radius of 10 km give typical Fermi energies of order $\SI{100}{\MeV}$, suggesting a significant muon abundance. 

In the absence of the gauged U(1)$_{\lmultau}$ symmetry, the muon abundance is determined purely by local chemical equilibrium between electrons and muons, namely $\mu_e(r)=\mu_\mu(r)$, where $\mu_\ell(r) \equiv \sqrt{m_\ell^2+(3\pi^2 n_\ell(r))^{2/3}}$ (see, for example, Ref.~\cite{Pearson:2018tkr}). Inclusion of the gauged U(1)$_{\lmultau}$ symmetry perturbs this picture inducing an unscreened electric field experienced by the muons. This leads to the following integral equation for the chemical potential: 
\begin{align} \label{eq:full_chem_pot}
\mu_e(r) = \mu_\mu(r) + g^{\prime 2}  \int_{\infty}^r \mathrm{d}r^\prime \frac{1}{r^{\prime 2}} \int_{0}^{r'} d r^{\prime\prime} r ^{\prime\prime 2 } n_\mu(r^{\prime\prime})
\end{align} 
This new potential term encompasses the additional energy cost of producing a muon, suppressing their production in the core of the NS. 
Given $n_e(r)$, one can then solve this equation for $n_\mu(r)$ to extract the muon number of an NS. 
In the limit where the potential term is negligible, the equation can be inverted to find $n_\mu(r)$ permitting a simple solution. This will be a good approximation when $g^{\prime2}N_\mu/4\pi r_\text{NS} \ll \mu_e $, where $ r _\text{NS}$ and $N _\mu$ are the radius and the total muon number of the NS, respectively. In the absence of an \lmultau force, $N_\mu$ is of order $ 1\%$ for typical NS parameters meaning for $g^\prime \gtrsim 10^{-18}$ one can no longer neglect the \lmultau force contribution to \cref{eq:full_chem_pot}. However, the muon abundance does not entirely disappear for larger gauge coupling. Muons will continue to be produced in the NS until the point at which the energy stored in the \lmultau field becomes of order the total energy stored in the electron gas, i.e.,
\begin{equation} 
\int_V \mathrm{d}V E ^2 _\mu \lesssim N _e \mu _e \quad\Longrightarrow\quad N _\mu \propto 1 / g ' \,,
\end{equation} 
where we have approximated the \lmultau electric field as $E_\mu \sim g^\prime N_\mu/r^2$, and $N_e$ is the total number of electrons in the NS. We therefore conclude that $ N_\mu $ is constant for $ g^\prime\lesssim  10^{-18}$ after which it then scales $\propto 1/g^\prime$. However, the observables of interest depend on the total muonic charge of the astrophysical objects, $g^\prime N_\mu$, which for $g^\prime > 10^{-18}$ will subsequently asymptote to a constant value. For the remainder of this paper, we will focus on the case where $g^\prime \lesssim 10^{-18}$, with a detailed derivation of the total muon number given in \cref{sec:muoncontent}, postponing the larger gauge coupling scenario to future work.

\section{Neutron star binary mergers}
\label{sec:inspiral_signal}

A new muonic force can have a dramatic effect on NS--NS and NS-BH binaries.   In the absence of exotic forces, the dynamics of these inspirals are determined by gravitational attraction and emission of gravitational waves.  Any new exotic force changes the dynamics in two different ways: $(i)$ the Yukawa force between the muon cores can accelerate or decelerate the merger, and $(ii)$ the emission of the mediator particle increases the energy loss of the system and accelerates the merger.  For concreteness, we consider a repulsive force mediated by a vector boson and follow the techniques advocated in~\cite{Alexander:2018qzg}.  In this section, we outline the technique via which we derive our new constraints. We postpone the detailed formulas to \cref{sec:app_waveform}.

If the muonic charges carried by the two astrophysical objects are denoted by $q_1$ and $q_2$, then the Yukawa force between them can be written as~\cite{Kopp:2018jom}
\begin{align}
\label{eq:Yukawa_force}
|{\bf F} (r)| = \dfrac{G_N m_1 m_2}{r^2} \left( 1+ \alpha \, e^{-\mv r} (1 + \mv r)\right)\,,
\end{align}
where $\alpha \equiv g^{\prime\,2} q_1 q_2/ (4\pi G_N m_1 m_2) = \tilde{q}_1 \tilde{q}_2 > 0$ and $r$ denotes the distance between the two astrophysical objects.  The masses of the NSs are denoted by $m_1$ and $m_2$, and the mass of the mediator vector boson is denoted by \mv.  The presence of such a new force modifies Kepler's law and the total energy of the system, $E_{\rm tot}$.  The energy loss rate of the system, $\diff E_{\rm tot}/ \diff t$, is then determined by the energy loss via gravitational waves, $\diff E_{\rm GW}/ \diff t$, and the energy loss via the emission of the new vector particle, $\diff E_{V}/ \diff t$,
\begin{align}
 \dfrac{\diff E_{\rm tot}}{\diff t} = -\frac{\diff }{\diff t} \left(E_{\rm GW} +E_V\right) \, ,
\label{eq:energy equation}
\end{align}
where $\diff E_{V}/ \diff t \propto \gamma$, and $\gamma \equiv g^{\prime\,2}  (q_1/m_1 - q_2/m_2)^2/(4\pi G_N)$ is the charge-to-mass ratio.  Due to the presence of this exotic force, both the plus and the cross polarizations of the GWs are affected.  We analytically calculate the GW amplitude and its phase to first order in $\alpha$ and $\gamma$.  We add post-Newtonian corrections following~\cite{Khan:2015jqa}. To derive the upper limits on $\alpha$ and $\gamma$ (and therefore $g^\prime$), we follow the standard Fisher information matrix analysis~\cite{Will:1994fb, Chamberlain:2017fjl, Vallisneri:2007ev, Porter:2015eha}. A complete prescription is given in \cref{sec:app_waveform}.

\section{Binary pulsars}
\label{sec:bin_pulsar}
Binary pulsars are a powerful probe of ultralight vectors. In a binary system, the motion of the pulsar and its companion are imprinted in the pulsar time-of-arrival data as an oscillation with a period, $ P_b $, which is typically of $ {\cal O} ( {\rm days} ) $. This is much larger than the pulsar period, which is $ {\cal O} ( {\rm msec} - 10~{\rm sec} ) $. While Keplerian orbits can explain the qualitative motion of a binary pulsar system, the precision of pulsar measurements allows the detection of deviations from classical mechanics due to relativistic effects. 

The deviations of a binary pulsar system from simple orbital motion can be described in terms of post-Keplerian parameters~\cite{1985AIHS...43..107D,1986AIHS...44..263D} (see also~\cite{Manchester:2015mda} for a review), the periastron precession, $ \dot{ \omega } $, the combination of gravitational redshift and Doppler shift, $ \gamma_d $, as well as the secular drop in the binary period, $ \dot{P} _b $, typically of $ {\cal O} ( 10 ^{ - 12} ) $. The three parameters depend on different combinations of the pulsar and companion masses. Since measurements of $ \dot{ \omega } $ and $ \gamma_d $ typically carry a much smaller uncertainty than $ \dot{ P} _b $, it is natural to use these to fix the masses and use $ \dot{P} _b $ to set constraints on new physics as advocated in Ref.~\cite{Krause:1994ar}.

For a binary pulsar system, the large muon abundance leads to the emission of the light \lmultau vector.\footnote{In addition, there are relativistic corrections to the force between two NSs which are prominent at slightly higher vector masses than radiation; however, these are subdominant to constraints from ensuring the new force does not eclipse gravity during NS mergers.} The subsequent rate of change in the energy of a pulsar relative to gravity is given by~\cite{Krause:1994ar}
\begin{equation} 
\frac{\big\langle\dot{E}_V \big\rangle}{\big\langle \dot{E}_\text{GW} \big\rangle} = \frac{5 \pi}{12}\,\gamma\,\frac{ g_V (\mv,\epsilon)}{g_\text{GR}(\epsilon)} \left(\frac{P_b }{ 2\pi G_N ( m _1 + m _2 )  }\right)^{2/3}\,,\label{eq:Eratio}
\end{equation}
where the elliptic correction functions for an eccentricity, $ \epsilon $, are:
\begin{align} 
g_V (\mv,\epsilon) & \equiv \sum _{n > n _0} 2 n^2 \left[{\cal J}_n ^{\prime\,2} (n \epsilon )  + \frac{ 1 - \epsilon ^2 }{ \epsilon ^2 } {\cal J}_n ^2 ( n \epsilon ) \right] \,,\notag \\ 
& \times \left\{ \sqrt{ 1 - \frac{  n _0 ^2 }{ n ^2 } } \left( 1 + \frac{1}{2} \frac{ n _0 ^2 }{ n ^2 } \right) \right\}  \\ 
g _{ {\rm GR}} ( \epsilon ) & \equiv  \frac{ 1 + ( 73/24) \epsilon ^2 + ( 37/96) \epsilon ^4 }{ ( 1 - \epsilon ^2 ) ^{ 7/2}}\,,
\end{align} 
where $ {\cal J}_n  $ is the Bessel function of $n$-th order and $ {\cal J}^\prime_n $ is its derivative, while the sum begins at $n _0 = \mv P_b /2\pi $. The observable for each binary pulsar is the ratio of the intrinsic change in the binary orbital period,\footnote{Note that $P_b^\text{int}$ is the observed value of the period drop which requires subtraction of the effects due to galactic rotation.} $P_b^\text{int}$, to the prediction from GR, $P_b^\text{GR}$.  Rewritten in terms of the energy ratios from above yields
\begin{equation} 
 \frac{ \dot{P}  _b ^{ {\rm GR}} }{ \dot{P}  _b ^{ {\rm int}} } = 1 - \frac{ \big\langle \dot{ E} _V \big\rangle }{  \big\langle \dot{ E} _{ {\rm GR}} \big\rangle } \,. 
\end{equation} 

We are now in a position to set constraints using current pulsar systems. There are many binary pulsars which have by now observed gravitational radiation including pulsar-(nonpulsating) NS binaries~\cite{Stairs:2002cw,Jacoby:2006dy,Ferdman:2014rna,vanLeeuwen:2014sca,Weisberg:2016jye}, a pulsar-pulsar binary~\cite{Kramer:2006nb}, pulsar-white dwarf binaries~\cite{vanStraten:2001zk,Bhat:2008ck,Freire:2012mg,Antoniadis:2013pzd}, as well as a pulsar-Oe-type star binary~\cite{Shannon2013}. In setting limits, different systems have different advantages. Since the energy ratio in \cref{eq:Eratio} is proportional to $ P_b^{2/3} $, large orbit binaries are ideal at probing low vector masses. On the other hand, binaries can only emit radiation efficiently if $ m _V \lesssim 2\pi / P _b $ leading to smaller binaries becoming effective at probing larger $ m _V $. Furthermore, we note that in order to have significant emission, the binary must carry a dipole moment, requiring the pulsar and its companion to have differing charge-to-mass ratios [as is apparent from the linear dependence of \cref{eq:Eratio} on the charge-to-mass ratio]. This quantity is maximized in pulsar-white dwarf or pulsar-visible star binaries where the charge of the white dwarf/star are negligible.

To set our constraints on gauged $ L _\mu - L _\tau $ using pulsar binaries we take the $ 2 \sigma $ limit on $ \dot{P}  _b ^{ {\rm int}} / \dot{P}  _b ^{ {\rm GR}} $ provided by experiments with the parameters summarized in \cref{fig:pulsarconst}
 of \cref{sec:app_pulsar_bin}. 

\section{Results and discussion}
\label{sec:discussion}

The constraints on a gauged U(1)$_{\lmultau}$ derived in \cref{sec:inspiral_signal,sec:bin_pulsar} are shown in \cref{fig:mV_vs_gV_constraints}. The lines marked NS--NS (NS--BH) merger, show the sensitivity that could be achieved with a dedicated LIGO/VIRGO analysis using GW170817 (S190814bv). The sensitivity curves in blue and black rely on dipole emission of the vector while the green curve arises from the new force between the two NSs. The right-hand boundaries of both these constraints are functions of $f_{\rm ISCO}$ (frequency of the innermost stable circular orbit). These boundaries lie at different \lmultau vector masses due to the different dependencies of the vector emission (force) on the angular velocity (binary separation) and therefore frequency of the binary. The dipole emission constraint asymptotes to a constant value at small vector masses as the only dependence on the vector mass arises in the step function [see Eq.~(B12) in the appendix]. Finally, the thin (thick) lines of a given colour indicate a pessimistic (optimistic) assumption on the muon abundance of the NS involved in the merger (additional details are given in \cref{sec:muoncontent}). 

The constraints shown in red are obtained using current binary pulsar data. We show the envelope of the constraints from all pulsars as the shaded purple region in~\cref{fig:mV_vs_gV_constraints}, while we defer the results for individual pulsars to \cref{fig:pulsarconst}
 of \cref{sec:app_pulsar_bin}. For eccentric binaries, the typical distance between the binaries changes significantly across the orbit and allows vector emission across different masses leading to the observed steplike pattern. Given the pulsar binary separations, the dipole radiation emission is only active for vector masses below ${\cal O}(\SI{E-18}{\eV})$. The sensitivity to the gauge coupling is weaker than the equivalent constraints from NS--NS mergers; however, they do not require additional analysis by LIGO and are a present constraint. As such the binary pulsar constraints serve as a robust alternative to LIGO's GW measurements. 

We now consider the validity of the constraints, which were derived assuming the muon number was unaffected by the presence of \lmultau. As argued in \cref{sec:muons_in_NSs} this approximation breaks down for $g^\prime \gtrsim 10^{-18} $, at which point $N_\mu \propto 1/g^\prime$. However, both the dipole and force effects depend on powers of the combination $g^\prime N_\mu$ which tends to a constant and therefore observable value for $g^\prime > 10^{-18}$. As the dipole emission for NS-NS binaries depends on $\gamma$ (the difference squared of the respective bodies charge-to-mass ratios), NS-NS binaries that are almost symmetric in charge-to-mass ratios are therefore unobservable irrespective of the gauge coupling size. This also explains the more pronounced difference in sensitivity between the optimistic and pessimistic muon abundances for the NS-NS dipole constraints in \cref{fig:mV_vs_gV_constraints} compared to either the force constraints or the NS-BH dipole constraints as only one astrophysical body contains muons.

\section{Conclusion and outlook}
\label{sec:conclusion}

We demonstrate the discovery reach to a new muonic force using neutron stars (NS) binary systems. The significant muon abundance inside NSs leads to two new effects: ($i$) dipole emission of the force mediator, and ($ii$) an additional force between binaries comprising of two NSs. These effects lead to changes in the dynamics of both inspiraling NSs and pulsar binaries. Measuring the deviation in the gravitational waveform and the pulsar period, respectively, are powerful tests of the presence of these new forces. Based on current data and assuming a U(1)$_{\lmultau}$ force as an example, \cref{fig:mV_vs_gV_constraints} shows the discovery reach to gauge couplings as small as $\mathcal{O}(\SI{E-20}{})$, orders of magnitude better than current probes. 

Given the estimated sensitivity of current gravitational wave experiments to these scenarios, we advocate for dedicated studies using improved calculations of the gravitational wave waveform confronted with data from the measured NS-NS merger event GW170817. In addition, future detection of BH--NS events, such as S190814bv, would allow for the isolation of the vector emission.

\section*{Acknowledgments}

We thank Iason Baldes, Kfir Blum, Raghuveer Garani, Admir Greljo, Edward Hardy, Joachim Kopp, Harikrishnan Ramani, Marko Simonovic and Yotam Soreq for useful discussions. In particular we would especially like to thank Evan McDonough for his hugely helpful correspondence. This work was initiated and performed in part at the Aspen Center for Physics, which is supported by National Science Foundation grant no.\,PHY1607611. T.O. received funding from the European Research Council under the European Union's Horizon 2020 research and innovation programme (grant agreement No.\ 637506, `$\nu$Directions') awarded to Joachim Kopp. J.D. is supported in part by the DOE under contract DE-AC02-05CH11231. 

{\it Note added:} During the preparation of this work, we became aware of~\cite{Poddar:2019wvu} which also considered using the muon content in NSs to set constraints on ultralight vectors using pulsar timing data.

\appendix

\section{Dependence on neutron star equation of state} 
\label{sec:muoncontent}

\begin{figure}
  \centering
  \includegraphics[width=\columnwidth]{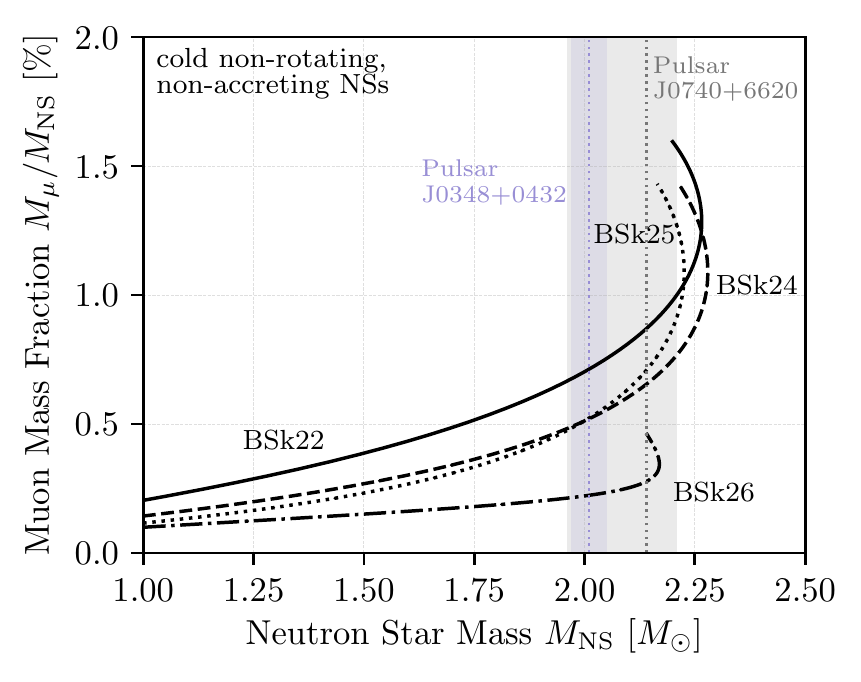}
  \caption{Muon content of a neutron star as a function of both the mass and EOS. For reference, the two heaviest measured neutron star masses from the NS--WD binary PSR J0348+0432 and the millisecond pulsar J0740+6620 \cite{2019arXiv190406759C} are shown in purple and gray, respectively.}
  \label{fig:muon_content}
\end{figure}

The exact muon content of a neutron star depends upon the QCD equation of state (EOS) relating the energy density to the pressure in the interior of the neutron star. We base our estimates of the muon content on the most recent Brussels-Montreal EOS~\cite{Pearson:2018tkr}, which is an update of older works based on the two- and three-nucleon force calculations of ~\cite{Potekhin:2013qqa,Pearson:2018tkr}. These EOS are all compatible with recent limits on the tidal deformability constraints from GW170817 \cite{TheLIGOScientific:2017qsa,Abbott:2018wiz}. Note, however, that two of these EOS, BSk22 and BSk26, are disfavoured due to neutron star cooling measurements. Measurements  suggest that relatively few neutron stars exhibit large cooling rates associated with the direct Urca process \cite{Page:2004fy,Gusakov:2004se}. The former EOS admits direct Urca processes for neutron star masses in excess of $1.151M_\odot$, and therefore anomalously large cooling rates in the majority of the neutron star population \cite{Klahn:2006ir}, while the latter EOS cannot support these processes at all which is in conflict with recent results in ~\cite{Brown:2017gxd}, suggesting the presence of this process in the neutron star MXB 1659-29. Nevertheless, we take the envelope (shown in \cref{fig:muon_content}) of all four EOS from ~\cite{Pearson:2018tkr} as a conservative estimate of the neutron star muon content as a function of the neutron star mass relevant for the different binary systems we consider. Finally, we note that similar EOSs are expected to hold for the description of rapidly rotating neutron star pulsars, while such rotating neutron stars in binary mergers are expected to aid in the extraction of the underlying EOS \cite{Harry:2018hke}. 

To estimate the muon content of the relevant neutron stars, we have utilized the fitting functions from~\cite{Pearson:2018tkr} for the relationship between the pressure and density of the neutron star. Based on these fitting functions for the EOS (see Eq.~(C4) in \cite{Pearson:2018tkr}), the Tolman-Oppenheimer-Volkoff (TOV) equations can be solved yielding a relation between the mass and density of the neutron star as a function of the radius. To determine the muon abundance, charge neutrality is assumed
\begin{align}
Y_p = Y_e + Y_\mu\,,
\end{align}
as well as equilibrium between the muons and electrons, resulting in chemical potentials that are the same $\mu_e = \mu_\mu$. Here $Y_i$ is the abundance $i=p,\,e,\,\mu$ of protons, electrons and muons defined as $Y_i\equiv n_i/n$ where $n$ is the total density and $n_i$ is the density of the species in question. From the chemical potential, under the assumption of degenerate electron and muon gases as well as sufficiently small \lmultau gauge coupling, we have
\begin{align}
m_e \left[1+\frac{(3\pi^2 n_e)^{2/3}}{m_e^2}\right]^{1/2} &= m_\mu \left[1+\frac{(3\pi^2 n_\mu)^{2/3}}{m_\mu^2}\right]^{1/2} \,,
\end{align}
where $m_e$ and $m_\mu$ are the electron and muon masses, respectively. This yields the muon number density 
\begin{align}
n_\mu &= \frac{m_e^3}{3\pi^2}\left[1+\frac{(3\pi^2 n_e)^{2/3}}{m_e^2}-\frac{m_\mu^2}{m_e^2}\right]^{3/2}\,.
\end{align}
To evaluate this expression we take the number density of electrons as a function of neutron density from Eq.~(C17) in \cite{Pearson:2018tkr}, allowing for the total mass of muons $M_\mu$ inside the neutron star to be determined for the different EOSs. The results of which are shown in \cref{fig:muon_content}. We observe that neutron stars with masses greater than a solar mass have $M_\mu \geq \SI{1.5E-3} M_\text{NS}$, while a two solar mass NS would have muon content in the range $ \SI{0.7E-3} M_\text{NS} \geq M_\mu \geq \SI{0.24E-3} M_\text{NS}$. For reference we also show the two heaviest observed neutron stars; the NS--WD binary PSR J0348+0432 and the millisecond pulsar J0740+6620 \cite{2019arXiv190406759C}.

\section{Modifications to the gravitational waveform}  
\label{sec:app_waveform}
\begin{figure*}
  \centering
  \includegraphics[width=0.495\textwidth]{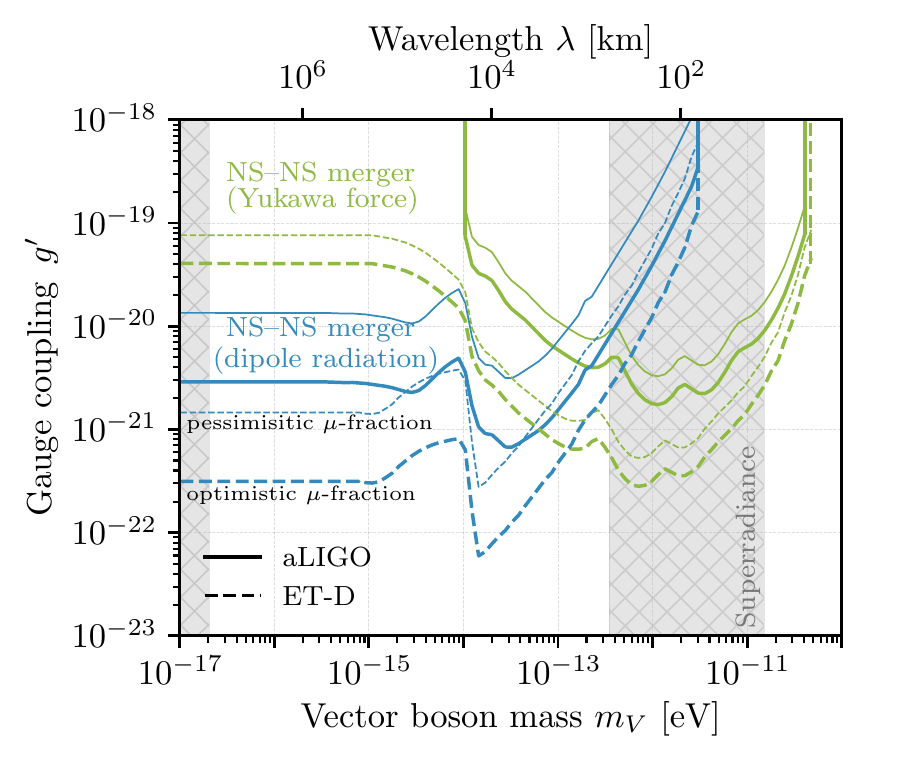}
    \includegraphics[width=0.495\textwidth]{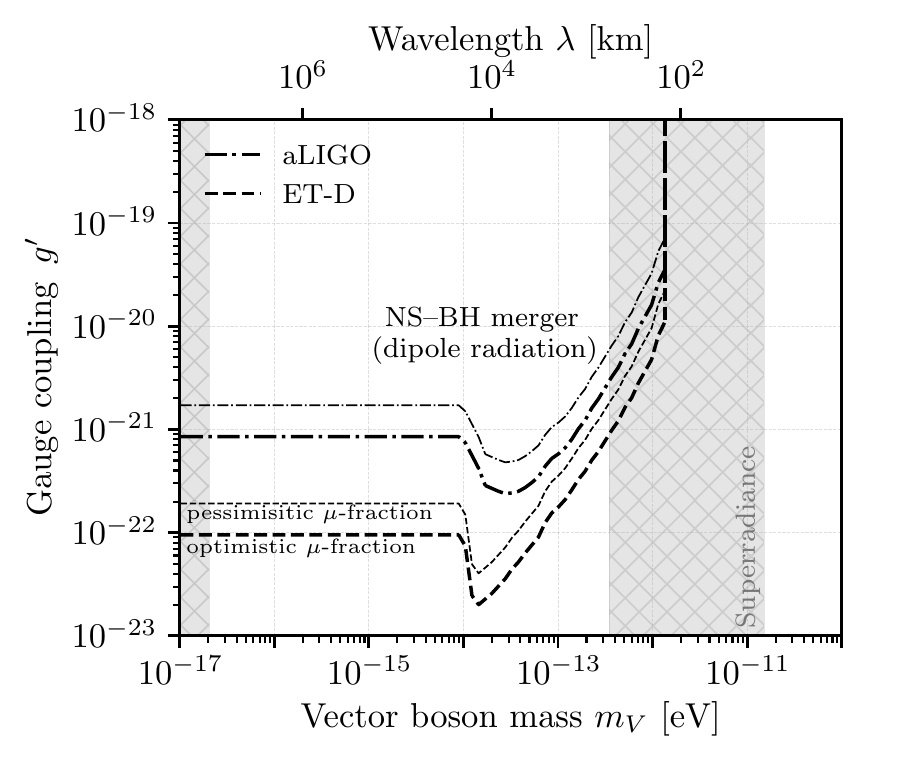}
  \caption{Constraints on the gauge coupling \gv as a function of the vector mass \mv for a gauged \lmultau symmetry. The regions above the curves indicate the projected parameter space where the presence of the light vector would lead to deviations incompatible with current measurements (solid lines) and future measurements (dashed and dot-dashed lines). {\bf Left}: projected constraints for an NS--NS event similar to GW170817. {\bf Right}: projected constraint assuming the observation of a NS--BH merger at a similar luminosity distance to GW170817.}
  \label{fig:mV_vs_gV_constraints_ET_projection}
\end{figure*}

Given the force between the two astrophysical objects as denoted by Eq.~(2), one can derive the orbital frequency, $\omega$, of the system,
\begin{eqnarray}
\omega^2 = \dfrac{G_N (m_1 + m_2)}{r^3}\, \left[1 - \alpha \, e^{- m_V r} (1 + m_V r) \right]\,,\phantom{xxxx}
\label{eq: orbital frequency}
\end{eqnarray}
where the negative sign before $\alpha$ denotes that the force is repulsive.  While writing the above equation, we neglect the spins of the astrophysical objects, and assume them to be point objects.  The orbital frequency is related to the frequency of the gravitational waves, $f_{\rm GW}$, as $f_{\rm GW} = \omega/\pi$.  The total energy of the system can be written as
\begin{align}
E_{\rm tot} &= - \dfrac{G_N m_1 m_2}{r} \, \left(1 - \alpha \, e^{-m_V r} \right) + \dfrac{1}{2}\mu r^2 \omega^2\,,
\label{eq: total energy}
\end{align}
where $\mu$ denotes the reduced mass of the system. The energy loss rate due to the emission of gravitational waves can be written as
\begin{eqnarray}
\dfrac{\diff E_{\rm GW}}{\diff t} = \dfrac{32}{5} G_N \mu^2 r^4 \omega^6 \,.
\label{eq: energy loss rate GW}
\end{eqnarray}
The energy loss due to the radiation of a light vector particle can be written as
\begin{eqnarray}
\dfrac{\diff E_V}{\diff t} &=& \dfrac{2}{3} \gamma \mu^2  \omega^4 r^2 \nonumber\\
&\times& {\rm Re} \left\{\sqrt{1- \left(\dfrac{m_V}{\omega}\right)^2} \left[1 + \dfrac{1}{2} \left(\dfrac{m_V}{\omega} \right)^2 \right] \right\} \,. \phantom{xxxx}
\label{eq: energy loss rate light vector particles}
\end{eqnarray}
where $\gamma$ is defined below Eq.~(3).

Due to the presence of this extra new force and a new way to lose energy from the system, the gravitational wave signature from the merger of two compact astrophysical objects change.  The two polarizations of the gravitational waves can be written as\,\cite{Alexander:2018qzg}
\begin{align}
h_+ (t) &= - \left(\frac{1+\cos^2 \iota}{2}\right)\mathcal{A}(t)\cos\left(2\phi_c + 2\phi\left(t-t_c\right) \right) \,,\nonumber\\
h_\times (t) &= -\left(\cos \iota\right) \mathcal{A}(t)\sin\left(2\phi_c + 2\phi\left(t-t_c\right)\right).
\label{eq: polarization of gravitational waves}
\end{align}
The inclination angle is denoted by $\iota$, and the time and phase of the system at coalescence is denoted by $t_c$ and $\phi_c$.  The orbital phase of the system is denoted by $\phi$ and we will mention how to calculate it later.  The amplitude of the signal is denoted by $\mathcal{A}(t)$ where
\begin{equation}
\mathcal{A}(t) = \frac{4 \, G_N}{D_L} \, \mu \,\omega^2(t) \, r^2(t)\,,
\label{eq: GW amplitude}
\end{equation}
where the luminosity distance to the source is denoted by $D_L$.  Given the detector responses, $F_+$ and $F_\times$, the strain detected by the detector is given by
\begin{align}
h(t) &= F_+ h_+ (t+t_c-t_0) + F_\times h_\times (t+t_c-t_0) \\
&= -\mathcal{A}(t+t_c-t_0)\bigg[\left(\frac{1+\cos^2\iota}{2}\right)F_+ \cos z\nonumber\\
&\qquad\qquad\qquad+ \left( \cos\iota \right) F_\times \sin z \bigg],
\end{align}
where $z=2 \left(\phi_c + \phi (t-t_0) \right)$ and the time in the detector frame when the coalescence is detected is given by $t_0$.  Defining $D_{\rm eff}$ and $\phi_0$ such that
\begin{eqnarray}
&& D_{\rm{eff}} = D_L \left[F_+^2\left(\frac{1+\cos^2\iota}{2}\right)^2 + F_\times^2 \cos^2\iota\right]^{-1/2} \,,\nonumber\\
&& \phi_0 = \phi_c - \arctan\left(\frac{2\cos\iota}{1+\cos^2\iota}\frac{F_\times}{F_+}\right),
\label{eq: Deff and phi0}
\end{eqnarray}
we can write the strain as
\begin{align}
h(t) = -\dfrac{4 \, \mu}{D_{\rm eff}} \omega^2 r^2\cos\left[2\phi_0 + 2\phi\left(t-t_0;m,\eta\right)\right]. \phantom{x}
\end{align}

In order to determine the upper limits on the parameters $\alpha$ and $\gamma$ from current observations, we need to determine the Fourier transform of $h(t)$.  Using the stationary phase approximation and restricting our calculations to first order in $\alpha$ and $\gamma$, we get\,\cite{Alexander:2018qzg}
\begin{widetext}
\begin{align}
\tilde{h}(f) &= -\sqrt{\dfrac{5\pi}{24}}  G_N^{5/6}  \dfrac{{\cal{M}}^2}{D_\text{eff}}(\pi {\cal{M}} f)^{-7/6}\left[\mathcal{A}_\text{PN} -\dfrac{\alpha}{3}C(x)
 - \dfrac{5\gamma}{96}(\pi  G_N	 m f)^{-2/3} \, \Theta\left(\frac{\pi f}{\mv} - 1\right)\right]e^{-i\Psi},
\label{eq: small coupling htilde}\\
\Psi &\equiv  2\pi f t_0 - 2\phi_0 - \dfrac{\pi}{4} + \dfrac{3}{128}\left(\pi  G_N {\cal{M}} f\right)^{-5/3}\left[\dfrac{20\alpha}{3}F_3\left(x\right) - \dfrac{5\gamma}{84}(\pi  G_N m f)^{-2/3} \, \Theta\left(\frac{\pi f}{\mv} - 1\right)\right]+\Psi_\text{PN},
\label{eq: Small couplings Phi}
\end{align}
\end{widetext}
where $x \equiv G_N^{1/3} m \, m_V (\pi m f)^{-2/3}$, $m\equiv m_1 + m_2$ is the reduced mass, and the chirp mass is denoted by $\mathcal{M} = \mu^{3/5} \, (m_1 + m_2)^{2/5}$.  The functions $F_3 (x)$ and $C(x)$ is defined as
\begin{align}
F_3(x) &=\left(\dfrac{180 + 180 x + 69 x^2 + 16 x^3 + 2 x^4}{x^4}\right)e^{-x} 
\nonumber \\ 
&+ \dfrac{21\sqrt{\pi}}{2x^{5/2}}{\rm erf}({\sqrt{x}}),  \label{eq: F3}\\
C(x) &= \left(1+x-2x^2\right) e^{-x}\,. 
\end{align}
where the error function is denoted by ${\rm erf}(x)$. Finally, the PN corrections from the gravity-only contribution take the form
\begin{align}
	\Psi_\text{PN} &= \sum_{n=0}^4 \frac{3}{128 \eta (\pi G_N m f)^{5/3}} \varphi_n (\pi G_N m f)^{n/3}\,, \\
	\mathcal{A}_\text{PN} &=  \sum_{n=0}^4 \mathcal{A}_n (\pi G_N m f)^{n/3}\,,
\end{align}
where $\eta \equiv m_1 m_2/m^2$ is the symmetric mass ratio and the coefficients of the sum are given in \cite{Khan:2015jqa}. We do not include corrections with $n > 4$ as these additional terms do not significantly alter the presented results.

In order to determine the upper limit on the new physics parameters, $\alpha$ and $\gamma$, we follow the Fisher information matrix analysis~\cite{Will:1994fb, Chamberlain:2017fjl, Vallisneri:2007ev, Porter:2015eha}. We denote the dimensionless spin parameters of the two astrophysical objects by $\chi_1$ and $\chi_2$.  We define the symmetric and antisymmetric dimensionless spin parameter as $\chi_s = (\chi_1 + \chi_2)/2$ and $\chi_a = (\chi_1 - \chi_2)/2$, respectively.  For the Fisher information matrix, we take the underlying parameters to be
\begin{align}
{\boldsymbol \theta} &= \left\{ \log{\cal A}, t_c, \phi_c, \log{\cal M}_c , \log\eta , \chi_s, \chi_a , \alpha, \gamma \right\}.
\label{eq: Fisher information matrix parameters}
\end{align}

\begin{figure*}
\begin{minipage}{.49\textwidth}
\resizebox{\columnwidth}{!}{
\def\arraystretch{1.3}
\setlength{\tabcolsep}{3pt}
\begin{tabular}{lcccccc}
\toprule[1.5pt]
Pulsar & $ P _b ( {\rm day} ) $  & $ \dot{P} _b ^{ {\rm int}} / \dot{P} ^{ {\rm GR}} _b  $  & $ \epsilon $    & $ M _1 [M _{\odot} ] $ & $ M _2[ M _{\odot} ] $ & Ref  \\ 
\midrule[.5pt] 
B1913+16(NS)& 0.323& $ 0.9983 \pm 0.0016$& 0.617 & 1.438& 1.390 & \cite{Weisberg:2016jye}\\ 
J0737-3039(P)&0.102 & $ 1.003 \pm 0.014 $  &0.088 & 1.3381& 1.2489&\cite{Kramer:2006nb} \\ 
J0437-4715(WD)&5.74 & $ 1.0 \pm 0.1 $& 0.00& 1.58 & 0.236& \cite{vanStraten:2001zk}\\ 
B1534+12(NS)& $ 0.421 $ & $ 0.91 \pm 0.06 $& 0.274 & 1.3452& 1.333& \cite{Stairs:2002cw}\\ 
B1259-63(O)& 1240&$  1.0 \pm 0.5 $ &0.870  &1.4 & 20 & \cite{Shannon2013}\\ 
J0348+0432(WD)& 0.102& $ 1.05 \pm 0.18 $  & 0.00& 2.01 & 0.172 &\cite{Antoniadis:2013pzd} \\ 
J1141-6545(WD)& 0.198& $ 1.04 \pm 0.06 $ &0.172 & 1.27 & 1.02 &\cite{Bhat:2008ck} \\ 
J1738+0333(WD)& 0.355& $ 0.94 \pm 0.13 $  & 0.00 & 1.46 & 0.19 &\cite{Freire:2012mg} \\ 
J1756-2251(NS)& 0.320& $ 1.08 \pm 0.03 $& 0.181  & 1.341 & 1.230 &\cite{Ferdman:2014rna} \\ 
J1906+0746(NS)&0.166 & $ 1.01 \pm 0.05$& 0.085 & 1.291&  1.322&\cite{vanLeeuwen:2014sca} \\ 
B2127+11C(NS)& 0.335& $ 1.00 \pm 0.03 $  &0.681 & 1.358& 1.354& \cite{Jacoby:2006dy}\\ 
\bottomrule[1.5pt]
\end{tabular}}
\end{minipage}
\begin{minipage}{.49\textwidth}
  \begin{center} 
\includegraphics[width=8.5cm]{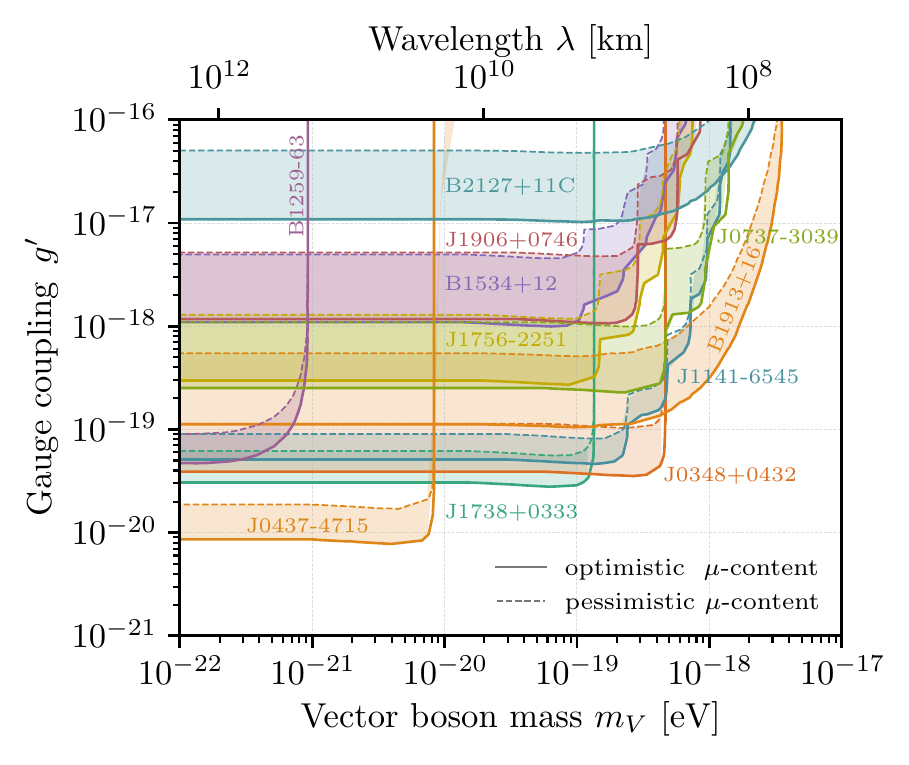} 
\end{center}
\end{minipage}
\caption{Data used to set the constraints on gauged \lmultau using pulsar binaries. {\bf Left}: the parameters for each pulsar relevant for computing the constraints. The type of companion star is shown with the name, denoting a nonpulsating neutron star by NS, pulsar by P, white dwarf by WD, and O as an Oe-type star. {\bf Right}: constraints from individual pulsars. For the parameter space above $g^\prime = \SI{E-18}{}$, several new effects as mentioned in Section V can become important.}
\label{fig:pulsarconst}
\end{figure*}

Using these 9 parameters, we can construct the 9 $\times$ 9 Fisher information matrix ${\bf \Gamma}$ whose components are given by
\begin{eqnarray}
\Gamma_{ab} \equiv \left( \frac{\partial h}{\partial \theta^a} \Bigg| \frac{ \partial h }{\partial \theta^b} \right) ,
\label{eq: Gamma matrix}
\end{eqnarray}
where $a$ and $b$ run from 1 to 9.  The 9 parameters, defined in \cref{eq: Fisher information matrix parameters}, are denoted by $\theta^{a, b}$.  The inner product is defined as
\begin{eqnarray}
\left( h_1 | h_2 \right) \equiv 4 \, {\rm Re}  \int _{f_{\rm low}} ^{f_{\rm high}} \frac{\tilde{h}_1 \tilde{h}_2^*}{S_n(f')} \mathrm{d}f' \,.
\label{eq: innerproduct}
\end{eqnarray}
We choose $f_{\rm low}$ following ~\cite{Alexander:2018qzg} and $f_{\rm high}$ = $f_{\rm ISCO}$, where $f_{\rm ISCO}$ denotes the frequency of the innermost stable circular orbit~\cite{Kopp:2018jom}. Finally, $S_n(f)$ denotes the spectral noise density of various gravitational wave detectors that we use in our analysis.  We use analytical forms of $S_n(f)$ as given in Ref.~\cite{Alexander:2018qzg} for the aLIGO and ET sensitivity curves.  The signal-to-noise ratio, $\rho$, is given by
\begin{eqnarray}
\rho \equiv \sqrt{\left(h | h \right)} \, .
\label{eq: SNR}
\end{eqnarray}
We define the covariance matrix, ${\bf \Sigma} \equiv {\bf \Gamma}^{-1}$.  The root-mean-squared error, that can be determined from an observation, for a given parameter $\theta^a$ is given by the square root of the $(a, a)$ component of ${\bf \Sigma}$.  The same expression also gives the 1$\sigma$ upper limits on the new physics parameters, $\alpha$ and $\gamma$,
\begin{eqnarray}
 \theta^a_{\rm new} \leq \sqrt{\Sigma^{\rm aa}} \, ,
\label{eq: Upper limit}
\end{eqnarray}
where  $\theta^a_{\rm new}$ denotes $\alpha$ and $\gamma$.

For the extraction of the $1\sigma$ upper limits we have chosen the parameter values to mimic the observed GW170817 event assuming slowly spinning neutron stars. This corresponds to the choices $m_1 = 1.46 M_\odot$, $m_2=1.27 M_\odot$ with $\chi_1 = 0.01$ and $\chi_2 = 0.02$. While for the effective luminosity distance we use $D_\text{eff} = \SI{40}{\mega \pc}$. In the left-hand panel of \cref{fig:mV_vs_gV_constraints_ET_projection} we show the resulting sensitivity curves for both aLIGO (solid lines) as well as the next generation ground based experiment ET (dashed lines). In much of the parameter space ET will improve sensitivity in $g^\prime$ by at least an order of magnitude. We note that the sensitivity curve and therefore resulting sensitivity for Cosmic Explorer is parametrically similar to ET. In the right-hand panel of \cref{fig:mV_vs_gV_constraints_ET_projection} we also show the projected sensitivity in the case of the observation of an NS--BH binary merger. Here we assume that $m_1=M_\text{BH} = 5 M_\odot$ and $m_2=M_\text{NS}=1.46 M_\odot$, again assuming small spins of both compact objects ($\chi_1 = 0.01$ and $\chi_2 = 0.02$). This type of merger is particularly sensitive as the charge-to-mass ratio is maximized given that that BHs carry zero charge under U(1)$_{\lmultau}$. Subsequently, the constraints are also less sensitive to the uncertainty in the muon content of the neutron star.

\section{Pulsar binary data}
\label{sec:app_pulsar_bin}
The data used to set the binary pulsar constraints is shown in \cref{fig:pulsarconst} ({\bf left}). For a given binary, the necessary parameters are the binary period, the change in the period relative to that predicted by gravity, the eccentricity, and the masses. We use the neutron star equation of state to compute the muon abundances of the neutron stars as described in the text. In \cref{fig:pulsarconst} ({\bf right}) we show the constraints from individual pulsar systems with optimistic and pessimistic assumptions on the muon abundance as described in \cref{sec:muoncontent}.

\bibliographystyle{JHEP}
\bibliography{refs}

\end{document}